# Solitary wave solution to the nonlinear evolution equation in cascaded quadratic media beyond the slowly varying envelope approximations

Amarendra K. Sarma[1*] and Anjan Biswas[2]

[1]Department of Physics, Indian Institute of Technology Guwahati, Guwahati-781039, Assam, India
[2]Department of Mathematical Sciences, Delaware State University, Dover, DE 19901-2277, USA
*E-mail: aksarma@iitg.ernet.in

We report exact bright and dark soliton solution to the nonlinear evolution equation derived by Moses and Wise [Phys. Rev. Lett. **97**, 073903, (2006)] for cascaded quadratic media beyond the slowly varying envelope approximations. The integrability aspects of the model are addressed. The traveling wave hypothesis as well as the ansatz method is employed to obtain an exact 1-soliton solution. Both bright and dark soliton solutions are obtained. The corresponding constraint conditions are obtained in order for the soliton solutions to exist.
**PACS number(s):** 42.65.Sf, 42.25.Bs, 42.65.Ky, 42.65.Re

Since the first experimental realization more than one decade ago, research in few cycle optical pulses has grown rapidly in recent years [1-2]. This tremendous boost is due to various reasons, primarily for richness in physics from both a fundamental and possible applications point of view in many diverse areas such as, ultrafast spectroscopy, metrology, medical diagnostics and imaging, optical communications, manipulation of chemical reactions and bond formation, material processing etc. [3-4]. Moreover, the field of light–matter interactions, high harmonic generation, extreme and single cycle nonlinear optics, and attosecond physics is also greatly influenced by the ultrashort pulses [5-9]. These pulses with duration of a few optical cycles are brief enough to resolve temporal dynamics on an atomic level like chemical reactions, molecular vibrations, and electron motion. They could be used for coherently exciting and controlling matter on a microscopic level since they are very broadband and can become extremely intense [10]. In this context, in order to describe the dynamics and propagation of few cycle optical pulses in nonlinear media, many authors looked for an appropriate mathematical model [11-13]. This is mainly owing to the fact that, because of the breakdown of the so called slowly varying envelope approximation (SVEA), the so called Nonlinear Schrodinger Equation (NLSE), which is routinely used as the governing equation for describing pulse propagation in a media, is inadequate in the few cycle regimes [2, 11]. In this context, the first widely accepted model has been developed by Brabec and Krausz [11]. Non-SVEA based mathematical models are also proposed by some authors [11-12]. Recently, owing to the efficient manipulation of spectral and temporal properties of few-cycle pulses through cascaded processes in quadratic materials, both theoretical and experimental research is getting tremendous boost in recent years [16-19]. And in this context, following the model proposed by Brabec and Krausz, Moses and Wise have derived a coupled propagation equations for ultrashort pulses in a degenerate three-wave mixing process in quadratic ($\chi^{(2)}$) media [20]. It may be noted that Moses-Wise model is restricted to the case of strongly mismatched interaction where the conversion efficiency to second or higher harmonics is negligible. Moses and Wise, based on their model, presented theoretical and experimental evidence of a new quadratic effect, namely the controllable self-steepening (SS) effect. The controllability of the SS effect is very useful in nonlinear propagation of ultrashort pulses as it may be used to cancel the propagation effects of group velocity mismatch. Recently, a modulation instability (MI) analysis of the Moses-Wise model is reported [21]. It is shown that subject to the fulfillment of the MI criterion and judicious choice of the parameters, MI could be generated in a cascaded-quadratic-cubic medium in both normal and anomalous dispersion regimes. As

MI could be considered as a precursor to soliton formation, this clearly motivates us to look Moses-Wise model a bit more closely. This work will shine light on the integrability aspects of the model. Moreover we will report exact 1- soliton solution to the propagation equation by using the traveling wave hypothesis [22] and the ansatz method [23]. Both bright and dark soliton solutions are discussed. It should be noted that both the travelling wave and ansatz methods are used quite extensively in obtaining solitary wave solutions to various nonlinear evolution equations such as Gross-Pitaeveskii equation [24], nonlinear Schrodinger equation with time dependent coefficients [25], higher order NLSE [26] and models related to propagation dynamics in proteins chains [27] etc.

The pulse propagation equation describing the ultrashort optical pulse propagation in quadratic nonlinear media beyond the slowly varying envelope approximation is given, in dimensionless units, as follows [21]:

$$i\frac{\partial u}{\partial z} + \alpha \frac{\partial^2 u}{\partial \tau^2} + \beta |u|^2 u = i\gamma u^2 \frac{\partial u^*}{\partial \tau} + i\delta |u|^2 \frac{\partial u}{\partial \tau} \tag{1}$$

Here in (1), $u(z,\tau)$ represents the wave profile. The coefficients $\alpha$ and $\beta$ refers to group velocity dispersion (GVD) and the self-phase modulation (SPM) respectively. Also, z and $\tau$ are the spatial and temporal variables respectively. The coefficient $\gamma$ is the self-steepening (SS) term that is induced by the group velocity mismatch (GVM), while $\delta$ is the so-called controllable self-steepening term, that originates from the slowly varying wave approximation [20]. This equation will be now solved by using two approaches, namely the traveling wave solution method and the so-called ansatz method. The first method will reveal a bright 1-soliton solution to (1) along with constraint conditions that needs to hold for the soliton solution to exist. The later method will give both a bright and a dark 1-soliton solution to (1) along with the necessary constraint conditions.

In order to integrate Eq. (1), by using the travelling wave solution method, we assume the following soliton solution with a permanent profile:

$$u(z,\tau) = g(z - v\tau)\exp[i(kz - \omega\tau + \theta)] \tag{2}$$

where $g(z - v\tau)$ is the amplitude portion while the remainder is the phase portion of the wave. $k, \omega, \theta$ and $v$ are respectively, the wave number, the frequency, the phase constant and the velocity of the wave. Putting (2) into Eq. (1) and then decomposing into real and imaginary parts, yield

$$\alpha v^2 g'' - (k + \alpha \omega^2)g + (\beta + \omega\gamma - \delta\omega)g^3 = 0 \tag{3}$$

and $(1 + 2\omega v\alpha) + (\gamma + \delta)vg^2 = 0$ \hfill (4)

where $g' = dg/ds$ and $g'' = d^2g/ds^2$ with $s = z - v\tau$. From Eq. (4) we obtain the following relations:

$v = -1/2\omega\alpha$ and $\gamma + \delta = 0$ \hfill (5)

These two relations could be considered as the constrains conditions for the solitons to exist where the first one gives the velocity of the soliton while the later relates the GVM induced SS and the controllable SS parameter. Now from Eq. (3), we obtain the following solution for $g$:

$$g = \lambda \operatorname{sech}\left[s\sqrt{(k + \alpha\omega^2)/\alpha v^2}\right], \text{ where } \lambda = \sqrt{2(k + \alpha\omega^2)/(\beta + \omega\gamma - \delta\omega)} \tag{6}$$

Hence, the bright 1-soliton solution of Eq. (1) is given by

$$u(z,\tau) = A\operatorname{sech}[B(z - v\tau)]\exp[i(kz - \omega\tau + \theta)] \tag{7}$$

where the amplitude of the soliton is given by $A = \lambda$ and the inverse width is given by $B = \sqrt{(k + \alpha\omega^2)/\alpha v^2}$. The relations for the amplitude and the width of the soliton introduce the constraints $(k+\alpha\omega^2)(\beta+\omega\gamma-\delta\omega)>0$ and $(k+\alpha\omega^2)\alpha>0$. These constrains along with the ones mentioned earlier must remain valid in order for the bright soliton solution to exist.

We will now integrate Eq. (1) to obtain both the bright and dark 1-soliton solution by using the ansatz method. It is worthwhile to mention that, following the study of MI of Moses-Wise model, with judicious choice of the GVM and the controllable SS parameter one may obtain either bright or dark soliton in a cascaded quadratic media. We assume,

$$u(z,\tau) = P(z,\tau)\exp[i\phi(z,\tau)] \tag{8}$$

where the amplitude and the phase component of the soliton are respectively given by $P(z,\tau)$ and $\phi(z,\tau)$ respectively. Here $\phi(z,\tau) = kz - \omega\tau + \theta$ as in the previous subsection. Substituting (7) in (1) and then decomposing into real and imaginary parts, we obtain:

$$-(k+\omega^2\alpha)P + (\beta+\omega\gamma-\delta\omega)P^3 + \alpha\frac{\partial^2 P}{\partial\tau^2} = 0 \tag{9}$$

and $\quad \dfrac{\partial P}{\partial z} - 2\omega\alpha\dfrac{\partial P}{\partial\tau} = (\gamma+\delta)P^2\dfrac{\partial P}{\partial\tau}$ \hfill (10)

Eq. (10) leads to the same velocity of the soliton given by Eq. (5) and the same constraint condition as in (5). We will now further analyse Eq. (9) to obtain bright and dark 1-soliton solution.
In order to obtain the bright soliton solution, let us assume,

$$P(z,\tau) = A\operatorname{sech}^p \xi \tag{11}$$

where $A$ is the amplitude of the soliton and $p$ is the unknown exponent. Also $\xi = B(z-v\tau)$, where $B$ is the inverse width of the soliton and $v$ is the velocity of the soliton. Thus, Eq. (9) simplifies to

$$-(k+\omega^2\alpha)\operatorname{sech}^p\xi + (\beta+\omega\gamma-\delta\omega)A^2\operatorname{sech}^{3p}\xi + \alpha B^2 v^2 p\left[p\operatorname{sech}^p\xi - (p+1)\operatorname{sech}^{p+2}\xi\right] = 0 \tag{12}$$

From (12), by the aid of balancing principle, equating the exponents $3p$ and $p+2$ leads to $p=1$. Then, setting the coefficients of the linearly independent functions $\operatorname{sech}^{p+j}$ for $j=0,2$ to zero leads to:

$$k = \alpha(B^2v^2 - \omega^2) \tag{13}$$

$$B = \sqrt{\frac{\beta+\omega(\gamma-\delta)}{2\alpha v^2}} A \tag{14}$$

which are the relations for the wave number and the amplitude-width. The second relation given by (14) introduces the constraint condition $\alpha(\beta+\omega\gamma-\delta\omega)>0$. Thus the bright 1-soliton solution to Eq. (1) as obtained by the ansatz method is given by (11), where the amplitude-width relation is given by (14) and the wave number is given by (13). The velocity of the soliton and the constraint condition is still given by Eq. (5). This method introduces a new constraint condition: $\alpha(\beta+\omega\gamma-\delta\omega)>0$ as a consequence of Eq. (14).

Now let us look for dark-soliton solution to Eq. (1). In order to obtain the dark 1-soliton solution to Eq. (1), we assume

$$P(z,\tau) = A\tanh^p \xi \tag{15}$$

In this case, the parameters $A$ and $B$ are referred to as the free parameters, while remaining parameters have the same interpretation. Adopting the similar procedure as above, we obtain:

$$k = -\alpha\left[\omega^2 + 2B^2v^2\right] \tag{16}$$

$$B = \sqrt{-\frac{\beta + \omega\gamma - \delta\omega}{2\alpha v^2}} A \qquad (17)$$

Eq. (17) shows that the constraint condition in this case is given by $\alpha(\beta + \omega\gamma - \delta\omega) < 0$. It is interesting to note that this condition is opposite to that of the bright soliton case. Hence the dark 1-soliton solution is given by $u(z,\tau) = A\tanh[B(z - v\tau)]\exp[kz - \omega\tau + \theta]$. The velocity is again given by Eq. (5). In order to check the stability of the both bright and dark soliton solution, we solve Eq. (1) by the so-called split-step Fourier method [28]. The parameters are chosen such that the respective constraints conditions are satisfied. In Fig.1 we depict the spatio-temporal evolution and the corresponding contour plots of the bright 1-soliton propagation. Fig. 2 depicts the contour plot of the dark 1-soliton propagation through the cascaded quadratic medium.

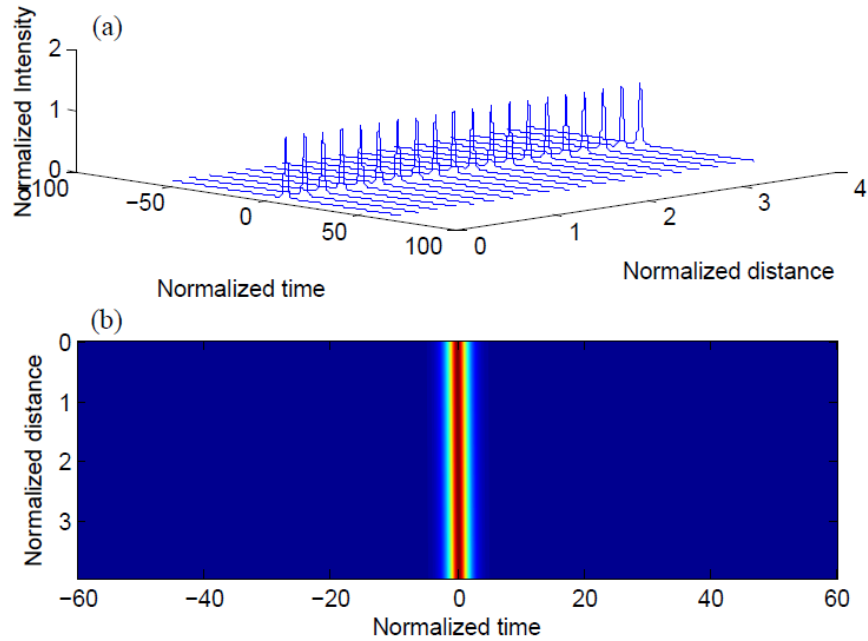

Fig. 1(a) Spatio-temporal evolution of bright-soliton propagation (b) the corresponding contour plot of (a). The parameters taken are: $\alpha = 1, \beta = 1, \gamma = -0.02, \delta = .02$ with A=1 and k=0.5.

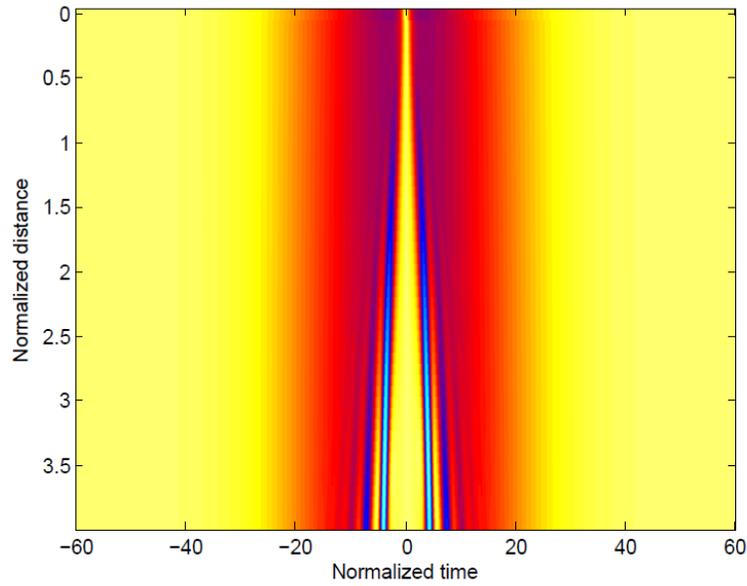

Fig. 2 Contour plot of dark soliton propagation. The parameters taken are: $\alpha=1, \beta=-1, \gamma=-0.02, \delta=.02$ with A=1 and k=0.5.

It could be observed that for the chosen parameters propagation of both bright and dark soliton is relatively stable. In fact, our numerical exploration shows that, both the bright and dark solitons are stable against its slight variation in amplitude during their propagation through the media. One can easily study the impact of various parameters on soliton propagation and draw useful conclusions, however in this report our objective is to report soliton solution and the constraints conditions under which bright or dark soliton may exist in cascaded quadratic medium. This study may enhance interest in exploring soliton phenomena in cascaded quadratic nonlinear media both theoretically and experimentally.

To conclude, we have reported exact bright and dark one-soliton solution to the nonlinear evolution equation derived by Moses and Wise for cascaded quadratic media beyond the slowly varying envelope approximations. The corresponding constraint conditions are obtained in order for the soliton solutions to exist. Numerical simulation indicates stability of the solitons subject to the judicious choice of parameters.